\documentclass{sig-alternate}

\usepackage{paralist}

\newcommand{\tagme}{\textsc{Tagme}}
\newcommand{\wikify}{\textsc{Wikify}}
\newcommand{\dc}{{\sc dc}}
\newcommand{\dt}{{\sc dt}}
\newcommand{\WikiDis}{{\sc Wiki-Disamb30}}
\newcommand{\WikiAnno}{{\sc Wiki-Annot30}}
\newcommand{\WikiLong}{{\sc Wiki-Long}}
\newcommand{\Pg}{P\!g}

\newenvironment{smallitemize}%
   {\vspace*{-4pt}
    \begin{itemize}\itemsep=0pt}%
   {\end{itemize}
    \vspace*{-2pt}}

\begin{document}

\title{Fast and accurate annotation of short texts \\ with Wikipedia pages}
\author{
	\alignauthor Paolo Ferragina \; and \; Ugo Scaiella\\
	\affaddr{Dipartimento di Informatica}\\
	\affaddr{University of Pisa, Italy}\\
	\email{\{ferragina, scaiella\}@di.unipi.it}
}


\maketitle



\begin{abstract}

We address the problem of cross-referencing text fragments with Wikipedia pages, in a way that synonymy and polysemy issues are resolved accurately and efficiently. We take inspiration from a recent flow of work \cite{cucerzan,soumen,wikify,witten}, and extend their scenario from the annotation of long documents to the annotation of short texts, such as snippets of search-engine results, tweets, news, blogs, etc.. These short and poorly composed texts pose new challenges in terms of efficiency and effectiveness of the annotation process, that we address by designing and engineering \tagme, the first system that performs an accurate and on-the-fly annotation of these short textual fragments. A large set of experiments shows that \tagme\ outperforms state-of-the-art algorithms when they are adapted to work on short texts and it results fast and competitive on long texts.

\end{abstract}

\section{Introduction}
\label{sec:intro}

The typical IR-approach to indexing, clustering, classification and retrieval, just to name a few, is that based on the bag-of-words paradigm. In recent years a good deal of work attempted to go beyond this paradigm with the goal of improving the search experience on (un-)structured or semi-structured textual data. In his invited talk at {\sc WSDM 2010}, S. Chakrabarti surveyed this work categorizing it in three main classes: (a) adding structure to unstructured data, (b) adding structure to answers, and (c) adding structure to queries while avoiding the complexity of elaborate query languages that demand extensive schema knowledge. In this paper we will be concerned with the first issue consisting of the identification of sequences of terms in the input text and their annotation with un-ambiguous entities drawn from a catalog. The choice of the catalog is obviously crucial for the success of the approach. Several systems nowadays adopt Wikipedia pages (or derived concepts) as entities, and implement the annotation process by hyper-linking meaningful sequences of terms with Wikipedia pages that are pertinent with the topics dealt by the input text. The choice of Wikipedia is dictated by the fact that the number of its pages is ever-expanding ($>3$ million English pages, and $>500$K pages in each major European language) and it offers the best trade-off between a catalog with a rigorous structure but with low coverage (like the one offered by the high-quality entity catalogs s.t. {\tt WordNet}, {\tt CYC}, {\tt OpenCYC}, {\tt TAP}), and a large text collection with wide coverage but unstructured and noised content (like the whole Web).

The first work that addressed the problem of annotating texts with hyper-links to Wikipedia pages was \wikify\ \cite{wikify}, followed by \cite{cucerzan}. Recently \cite{soumen,witten} yielded considerable improvements by proposing several new algorithmic ingredients, the most notable ones are: (i) a measure of {\em relatedness} among Wikipedia pages based on the overlap of their in-links; and (ii) the modeling of the process as the search for the annotations that {\em maximize} some global score which depends on their coherence/relatedness and other statistics. These ingredients allowed Milne\&Witten \cite{witten} to achieve an F-measure of 74.8\% over long and focused input texts, and Chakrabarti et al \cite{soumen} to improve the recall by still obtaining comparable precision.

In this paper we add to this flow of work the specialty that the input texts to be annotated are {\em very short}, namely, they are composed of few tens of terms. The context of use we have in mind is the annotation of either the snippets of search-engine results, or the tweets of a Twitter channel, or the items of a news feed, or the posts of a blog, etc.. As an example, let us consider the following (recent) news: ``Diego Maradona won against Mexico''. Our goal is to detect ``Diego Maradona'' and ``Mexico'' as meaningful term sequences to be annotated (hereafter called {\em spots}), and then hyper-link them with the senses represented by the Wikipedia pages which deal with the soccer player (now Argentina's coach) Maradona and the football team of Mexico. The key difficulty of this process is to detect on-the-fly those spots and their pertinent senses among the (possibly many) Wikipedia pages that are linked by those anchors in Wikipedia. In fact ``Diego Maradona'' is the anchor of one Wikipedia page, whereas Mexico is the anchor of $154$ Wikipedia pages. A good annotator should therefore be able to {\em disambiguate} ``Mexico'' by linking this spot with the page dealing with the football team, rather than the state. Furthermore, it should {\em prune} the other spots ``won'' and ``against'' which are actually anchors in Wikipedia but obviously result not meaningful in the present news.

It is easy to argue that these poorly composed texts pose new challenges in terms of efficiency and effectiveness of the annotation process, which (1) should occur on-the-fly, because in those contexts data may be retrieved at query time and thus cannot be pre-processed, and (2) should be designed properly, because the input texts are so short that it is difficult to mine significant statistics that are rather available when texts are long. The systems of \cite{soumen,witten} are designed to deal with reasonably long texts, and indeed  they either depend on statistics that hinge on many well-focused spots \cite{witten} or they compute sophisticated scoring functions that make the whole process slow \cite{soumen}.

Given these limitations we have designed and implemented \tagme\ the first software system that, on-the-fly and with high precision/recall, annotates short texts with pertinent hyper-links to Wikipedia pages. \tagme\ uses as spots (to be annotated) the sequences of terms composing the anchor texts which occur in the Wikipedia pages, and it uses as  possible senses for each spot the (possibly many) pages pointed in Wikipedia by that spot/anchor. \tagme\ resolves synonymy and polysemy issues among the potentially many available mappings (spot-to-page) by finding a collective agreement among them via {\em new} scoring functions which are {\em fast} to be computed and {\em accurate} in the finally produced annotation. This paper will detail the algorithmic anatomy of \tagme\ and present a large and variegate set of experiments that will validate the algorithmic choices made in the design of \tagme, and will experimentally compare \tagme\ against the two best-known systems \cite{soumen,witten} both on short and long texts. Sect. \ref{sub:results} will detail our achievements, here we summarize that on short texts \tagme\ outperforms the best known systems either in accuracy or speed, or both. On long texts, \tagme\  is competitive in accuracy and still very fast, with a time complexity that grows {\em linearly} with the number of processed anchors (cfr. \cite{soumen}'s quadratic time complexity). \tagme\ is available for test at {\tt http://tagme.di.unipi.it}.\footnote{A preliminary (and short) description of \tagme\ has been published as poster in CIKM 2010 \cite{tagme}. The present paper contains an engineered version of \tagme\ with a larger set of experiments, comparisons and findings.}

\section{Notation and terminology}
\label{sec:notation}

A \emph{(text) anchor} for a Wikipedia page $p$ is a text used in another Wikipedia page to point to $p$. In Wikipedia, this can be the title of $p$, one of its synonyms or acronyms, or even it may consists of a long phrase which might be (much) different syntactically from $p$'s title. As an example, an anchor text for the page ``Nintendo DS'' is the acronym ``nds'' as well as the phrase ``Gameboy ds'' or ``Nintendo Dual Screen''. For coverage purposes, we enrich all anchors of $p$ with the title of the redirect pages that link to $p$. This approach derives a total of about 8M distinct anchors from Wikipedia (English). In this paper, following \cite{soumen}, we will interchangeably use the terms {\em anchor} and {\em spot}.

Because of polysemy and variant names, the same anchor $a$ may occur in Wikipedia many times pointing to many different pages. We denote this set by $\Pg(a)$, and use the notation: $freq(a)$ to denote the number of times $a$ occurs in Wikipedia (as an anchor or not); and $link(a)$ to indicate the number of times the text $a$ occurs as an anchor in Wikipedia (of course $link(a) \leq freq(a)$). Also we use $lp(a) = link(a)/freq(a)$ to denote the \emph{link}-probability that an occurrence of $a$ is an anchor pointing to some Wikipedia page; and use $\Pr(p|a)$ to denote the \emph{prior}-probability that an occurrence of an anchor $a$ points to a specific page $p \in \Pg(a)$. This latter is also called \emph{commonness} of $p$.

The annotation of an anchor $a$ with some page $p \in \Pg(a)$ is denoted by $a \mapsto p$. Often $a$ has more senses, thus $|\Pg(a)|>1$, so we call {\em disambiguation} the process of selecting one of the possible senses of $a$ from $\Pg(a)$. It goes without saying that not all occurrences of the anchor $a$ should be considered as meaningful and thus be annotated. So we follow \cite{soumen} and introduce a {\em fake page} $\textsc{na}$ that is used to {\em prune} the un-meaningful annotations, via the dummy mapping $a \mapsto \textsc{na}$.

\section{Related Works}
\label{sec:related}

The literature offers two main approaches to enrich a (possibly short) text with additional structure and information that may empower subsequent IR-steps such as clustering, classification, or mining.

One approach consists of extending the classic term-based vector-space model with additional  dimensions corresponding to features (concepts) extracted from an external knowledge base, such as {\sc DMOZ}  \cite{snaket, esa-old}, Wikipedia \cite{gabrilovich,clustering-short-text2,clustering-short-text}, or even the whole Web (such as the Google's kernel \cite{g-kernel}). Probably the best achievements have been obtained by querying Wikipedia (titles or entire pages) by means of short phrases (possibly single terms) extracted from the input text to be contextualized. The result pages (typically restricted to the top-$k$) and their scores (typically tf-idf) are used to build a vector that is considered the ``semantic representation'' of the phrase and is finally used in classification \cite{gabrilovich}, clustering \cite{clustering-short-text2,clustering-short-text}, or searching \cite{query-by-doc} processes. The pro of this approach is to extend the bag-of-words scheme with more concepts, thus possibly allowing the identification of {\em related} texts which are syntactically far apart. The cons resides in the contamination of these vectors by un-related (but common) concepts retrieved via the syntactic queries.

In order to overcome these limitations, some authors have tried to annotate only the {\em salient} text fragments present in an input text, without resorting to the vector-space model. Their key idea is to identify in the input text short-and-meaningful sequences of terms and connect them to unambiguous senses drawn from a catalog. The catalog can be formed by either a small set of specifically recognized types, most often People and Locations  (aka Named Entities), or it can consists of millions of senses drawn from a large knowledge base, such as Wikipedia. In the former case (see e.g. \cite{google-ne,semawiki}), substantial training and/or human effort is required to eventually produce a ``coarse'' annotation: these systems would probably recognize that a sequence of terms, say {\em Michael Jordan}, is the name of a person, but they would miss to disambiguate which person the occurrence is referring to (in fact, Wikipedia contains 7 persons with the name {\em Michael Jordan}). In the latter case, the annotation can take advantage of million senses (currently more than 3 million English pages, and more than 500K pages in each major European language) and several million relations among them. This catalog is ever-expanding and currently offers the best trade-off between a catalog with a rigorous structure but with low coverage (like the one offered by the high-quality entity catalogs s.t. {\tt WordNet}, {\tt CYC},  {\tt TAP}), and a large text collection with wide coverage but unstructured and noised content (like the whole Web).

To our knowledge the first work that deployed this huge knowledge base of senses and relations to efficiently and accurately cross-reference long documents was \wikify\ \cite{wikify}, soon followed by \cite{cucerzan}. Recently, Milne\&Witten \cite{witten} proposed an approach that yielded considerable improvements by hinging on three main ingredients: (i) the identification in the input text of a set $C$ of so called \emph{context pages}, namely Wikipedia pages pointed by anchors that are not ambiguous (because they link to just one page/sense); (ii) a measure $rel(p_1,p_2)$ of {\em relatedness} between two pages $p_1,p_2$ based on the overlap between their in-linking pages in Wikipedia; and finally (iii) a notion of \emph{coherence} between a page $p$ and the other context pages in $C$. Given these, the disambiguation of an anchor $a$ was obtained by using a classifier that exploited for each sense $p \in \Pg(a)$: the commonness $\Pr(p|a)$ of the annotation $a \mapsto p$, the relatedness $rel(p,c)$ between the candidate sense $p$ and all other context pages $c \in C$ (which are un-ambiguous), and the coherence of each $c$ with respect to the entire input text. Then anchor pruning was performed by using another classifier that mainly exploited the location and the frequency of $a$ in the input text, its link probability, the relatedness between $p$ and the un-ambiguous pages of $C$, and the confidence of disambiguation assigned by the classifier to $a \mapsto p$ (at the previous step). In \cite{witten} the authors showed an F-measure of 74.8\% but this holds for {\em ``reasonably long and focused''} texts,\footnote{This is the response message of the system, available at \texttt{http://wikipedia-miner.sourceforge.net}, when the input text is too short.} which seems unsuitable for our scenario in which we need to process short (and thus potentially ambiguous) input texts.

Last year, Chakrabarti and his group \cite{soumen} proposed an annotator based on two other novelties. The first one was to evaluate an annotation $a\mapsto p$ with two scores: one local to the occurrence of $a$ (and involving $12$ features) and the other global to the entire input text (and involving all the other annotations and a relatedness function inspired by \cite{witten-rel}). The second novelty was to model the entire annotation process as a search for the mapping that {\em maximizes} some global score, via the solution of a (sophisticated) quadratic assignment problem. Extensive experiments showed that \cite{soumen}'s approach yields precision comparable to Milne\&Witten's system but with a considerable higher recall. Unfortunately, the system is slow since it takes $>2$ seconds to annotate a text of about $15$ anchors (see Figure 13 of \cite{soumen}); this is due to the sophisticated annotation process (recall the quadratic-assignment problem above) and the many term-vector comparisons and computations. This annotation speed is acceptable for an off-line setting, like the one considered in \cite{soumen}, but it is unsuitable for our setting where we wish to annotate \emph{on-the-fly} many short texts (possibly coming from the results of a search engine, or a tweet channel).

\smallskip
In summary, the systems of \cite{soumen,witten} seem unsuitable to annotate on-the-fly short and poorly composed texts, given that they either depend on statistics that hinge on many well-focused spots \cite{witten} or they compute sophisticated scoring functions that make the whole process slow \cite{soumen}. The numerous experiments in Sect. \ref{sec:evaluation} will sustain this intuition, and will validate the use of \tagme\ not only for the annotation of short texts but also for the long ones.

\subsection{Our Results}
\label{sub:results}

The first goal of this paper is to describe the algorithmic anatomy of \tagme, the first software system that annotates short text fragments on-the-fly and with high precision/recall by cross-referencing meaningful text spots (i.e. anchors drawn from Wikipedia) detected in the input text with one pertinent sense (i.e. Wikipedia page) for each of them. This annotation is obtained via two main phases, which are called anchor disambiguation and anchor pruning. Disambiguation will be based on finding the {\em ``best agreement''} among  the senses assigned to the anchors detected in the input text. Pruning will aim at discarding the anchor-to-sense annotations which result {\em not pertinent} with the topics the input text talks about. So the structure of \tagme\ mimics the one of \cite{soumen,witten}'s systems but introduces some new scoring functions which improve the speed and accuracy of the disambiguation and pruning phases. The algorithmic contribution of \tagme's design therefore consists of:

\begin{smallitemize}

\item a new voting scheme for anchor disambiguation that builds upon the relatedness function proposed in \cite{witten-rel} to find the collective agreement among all anchor-sense matches detected in the (short) input text. The specialty of this voting is that it is simple and thus fast to be computed (cfr. \cite{soumen}'s quadratic assignment problem), and it judiciously combines the relatedness among {\em all} candidate annotated-senses (cfr. \cite{witten}'s un-ambiguous pages only) in order to account for the sparseness of the anchors which is typical of the annotation of short texts.

\item the design and test of several pruning schemes which build upon two simple features extracted from each candidate annotation: the link probability of its anchor and the {\em coherence} between that annotation and {\em all} other candidate annotations detected in the input text. Although these features have been already used by \cite{soumen,witten}, \tagme\ will combine them in many new ways. The final result will be a large spectrum of possible pruning approaches, from which we will choose the final \tagme's pruner that will consistently improve known systems, yet remaining sufficiently simple and thus fast to be computed.

\end{smallitemize}

\smallskip The second goal of this paper is the execution of a large and variegate set of experiments on publicly available datasets \cite{soumen} as well as on new large datasets that we have created and made available to the community.\footnote{{\tt http://acube.di.unipi.it/datasets}} These experiments will aim at (i) testing the novel disambiguation and pruning algorithms described above, in order to set the best choice for \tagme, (ii) comparing \tagme\ against the two best-known systems--- namely, Chakrabarti's and Milne\&Witten's--- both on short and long texts, in order to derive principled conclusions about speed and accuracy issues of these annotators. In this respect, the contribution of our paper will be the following:

\begin{smallitemize}

\item On short texts we will show that \tagme\ outperforms Milne\&Witten's system by yielding an F-measure of about 78\% (versus their 69\%), with the possibility to balance precision (up to 90\%) vs recall (up to 80\%), at similar annotation speed. The system of Chakrabarti et al has not been tested because unavailable\footnote{S. Chakrabarti's personal communication.}; anyway, as commented in Sect. \ref{sec:related}, it could not be used in our context because it is very slow since it takes $>2$ secs per $15$ anchors \cite{soumen}. This is more than one order of magnitude slower than \tagme, which takes less than 2ms per anchor (see Sect. \ref{sub:time} for details).

\item On long texts, \tagme\ is competitive with the two above systems in terms of accuracy with the advantage of offering a faster speed (still less than 2ms per anchor). This is due to its algorithmic structure that guarantees a time complexity {\em linear} in the number of processed anchors (cfr. \cite{soumen}'s quadratic time complexity), and an efficient internal-memory utilization bounded by 200Mb (cfr. \cite{witten}'s software that uses more than 1.5Gb).

\end{smallitemize}

\section{The anatomy of TAGME}
\label{sec:tagme}

\tagme\ indexes some distilled information drawn from the Wikipedia snapshot of November 6, 2009.

\smallskip {\em Anchor dictionary.} We took all anchors present in the Wikipedia pages, and augmented them with the titles of redirect pages plus some variants of the page titles, as suggested in \cite{cucerzan}. We then removed the anchors composed by one character or just numbers, and also discarded all anchors $a$ whose absolute frequency ($link(a)<2$) or its relative frequency ($lp(a) < 0.1\%$) was small enough that we could argue $a$ being unsuitable for annotation and probably misleading for disambiguation. The final dictionary contains about 3M anchors, and it is indexed by Lucene\footnote{\tt http://lucene.apache.org}.
	
\smallskip {\em Page catalog.} We took all Wikipedia pages and discarded disambiguation pages, list pages, and redirect pages, because un-suitable as senses for anchor annotation. The remaining 2.7M pages were indexed by Lucene.

\smallskip {\em In-link graph.} This is a directed graph whose vertices are the pages in the Page Catalog, and whose edges are the links among these pages derived from the Wikipedia-dump called ``Page-to-page link records''. This graph contains about $147$M edges, and is indexed in internal-memory by Webgraph\footnote{\tt http://webgraph.dsi.unimi.it}.

\smallskip
\tagme\ uses these data structures to annotate a short text via three main steps: (anchor) parsing, disambiguation and pruning. Parsing detects the anchors in the input text by searching for multi-word sequences in the Anchor Dictionary; Disambiguation judiciously cross-references each of these anchors with one pertinent sense drawn from the Page catalog; Pruning discards possibly some of these annotations if they are considered not meaningful for contextualizing the input text. Everything is designed to occur on-the-fly and achieve high precision/recall. Details follow.

\subsection{Anchor parsing}
\label{sub:anchorParsing}

\tagme\ receives a short text in input, tokenizes it, and then detects the anchors by querying the \emph{Anchor dictionary} for sequences of up to $6$ words. Since anchors may overlap or be substring one of another, we need to detect their boundaries. We simplified the approach of \cite{cucerzan} in the following way: if we have two anchors $a_1,a_2$ s.t. $a_1$ is a (word-based) substring of $a_2$, we drop $a_1$ only if $lp(a_1) < lp(a_2)$. This is because $a_1$ is typically more ambiguous than $a_2$ (being one of its substrings), and editors like to link more specific (longer) word sequences. Therefore, we prefer to discard $a_1$ in order to ease the subsequent disambiguation task. As an example, consider $a_1 =$ ``jaguar'' and $a_2 = $ ``jaguar cars'': in this case if we didn't discard $a_1$, disambiguation task would uselessly handle all possible senses of ``jaguar'' thus slowing down the process and making it more cumbersome.

On the other hand, it might be the case that $lp(a_1) > lp(a_2)$. Given that $freq(a_1)\geq freq(a_2)$, this may occur only if $link(a_1) \gg link(a_2)$. This is the case when $a_2$ adds a non-meaningful word to $a_1$ that nonetheless identifies some senses. As an example, consider $a_1 = $ ``act'' and $a_2 = $ ``the act'' for which it is $lp(a_1) > lp(a_2)$: in fact, ``act'' refers to a huge amount of possible senses (Act of parliament, Australian Capital Territory, Act of a drama, Group Action, etc.), while ``the act'' is the name of a band and the title of a musical with a consequent small number of link occurrences. In this case we keep both anchors because, at this initial step of the annotation process, we are not able to make a principled pruning.

\subsection{Anchor disambiguation}
\label{sub:disambig}

This phase takes inspiration from \cite{soumen,witten-rel,witten}, but extends their approaches to work accurately and on-the-fly over short texts. As in \cite{soumen}, we aim for the collective agreement among all senses associated to the anchors detected in the input text and, as in \cite{witten}, we take advantage of the un-ambiguous anchors (if any) to boost the selection of these senses for the ambiguous anchors. However,
unlike these approaches, we propose new disambiguation scores that are much simpler, and thus faster to be computed, and take into account the sparseness of the anchors and the possible lack of un-ambiguous anchors in short texts.

More precisely, given a set of anchors ${\cal A}_T$, detected in the short input fragment $T$, \tagme\ tries to disambiguate each anchor $a \in {\cal A}_T$ by computing a score for each possible sense $p_a$ of $a$ (hence $p_a \in \Pg(a)$). This score is based on a new notion of ``collective agreement'' between the sense $p_a$ (Wikipedia page) and the senses (pages) of all other anchors detected in $T$. The (agreement) score of $a\mapsto p_a$ is evaluated by means of a {\em voting scheme} that computes for each other anchor $b \in {\cal A}_T \setminus \{a\}$ its {\em vote} to that annotation. Given that $b$ may have many senses (i.e. $|\Pg(b)| > 1$) we compute this vote as the {\em average relatedness} between each sense $p_b$ of $b$ and the sense $p_a$ we wish to associate to $a$. The relatedness between the two Wikipedia pages $p_a$ and $p_b$ is computed as suggested in \cite{witten-rel} as:

$$rel(p_a,p_b) = \frac{ \log(\max(|in(p_a), in(p_b)|)) - \log(|in(p_a) \cap in(p_b)|)}{\log(W) - \log(\min(|in(p_a), in(p_b)|))}$$

where $in(p)$ is the set of Wikipedia pages pointing to page $p$ and $W$ is the number of pages in Wikipedia. Hence the voting given by anchor $b$ to the annotation $a\mapsto p_a$ is:

\[
vote_b(p_a) =
	\frac{
		\sum_{p_b \in \Pg(b)} {rel(p_b,p_a) \cdot \Pr(p_b|b)}
	}{
		|\Pg(b)|
	}
\]

We notice that the average is computed by weighting each relatedness $rel(p_a,p_b)$ with the commonness of the sense $p_b$ (i.e. $\Pr(p_b|b)$), because we argue that not all possible senses of $b$ have the same (statistical) significance. So if $b$ is un-ambiguous, it is $\Pr(p_b|b)=1$ and $|\Pg(b)|=1$, and thus we have $vote_b(p_a) = rel(p_b,p_a)$ and hence we fully deploy the unique senses of the un-ambiguous anchors (as it occurred in \cite{witten}). But if $b$ is polysemous, only the senses $p_b$ related to $p_a$ will mainly affect $vote_b(p_a)$ because of the use of the relatedness score $rel(p_b,p_a)$.

Finally, the total score for the annotation $a\mapsto p_a$ is computed as the sum of the votes given by all other anchors $b$ detected in $T$:

$$rel_a(p_a)= \sum_{b \in {\cal A}_T \setminus \{a\}} {vote_b(p_a)}$$

This score is not enough to obtain an accurate disambiguation, so we combine it with the commonness of the sense $p_a$ for $a$ (i.e. $\Pr(p_a|a)$), used as the ``statistical support'' for the significance of this annotation. There are of course many possible ways to combine these two values. In this paper we investigate two approaches:  Disambiguation by Classifier (shortly \dc) and Disambiguation by Threshold (shortly \dt). \dc\ uses a classifier that takes the above two scores as features and computes a value that can be interpreted as the {\em ``probability of correct disambiguation''} for the mapping $a\mapsto p_a$.
Then it annotates $a$ with the sense $p_a \in \Pg(a)$ that reports the highest classification score.

On the other hand, \dt\ avoids the use of classifiers and recognizes a roughness in the value of the voting-score $rel_a(p_a)$ among all $p_a \in \Pg(a)$. So it first determines the sense $p_{best}$ that achieves the highest relatedness $rel_a(p_{best})$ with the anchor $a$, and then identifies the set of other senses in $\Pg(a)$ that yield about
the same value of $rel_a(p_{best})$, according to some fixed threshold $\epsilon$. Finally \dt\
annotates $a$ with the sense $p_a$ that obtains the {\em highest commonness} $\Pr(p_a|a)$ among
these {\em top-$\epsilon$} senses.

Given that speed is a main concern, both \dc\ and \dt\ discard from the above computation all senses whose commonness is lower than a properly set threshold $\tau$. In fact, as illustrated in \cite{witten}, the distribution of $\Pr(p|a)$ follows a power law so we can safely discard pages at the tail of that distribution. The setting of $\tau$ clearly affects the precision of the disambiguation process: if $\tau$ is too large, precision decreases because we would discard many pertinent senses; if $\tau$ is too small, speed and recall decrease. In Sect.~\ref{sub:eval-disamb} we will perform a wide set of experiments to evaluate these two algorithms and their parameter settings.

\smallskip We conclude this section by pointing out the {\em key differences} between \tagme's and the disambiguation-scores proposed by Milne\&Witten and Chakrabarti et al  (see Sect.~\ref{sec:related}). As for the former, disambiguation was based only on un-ambiguous anchors which are possibly missing in the short input texts. The consequence on the performance of \tagme's disambiguation, as reported in the following Table \ref{tab:disambALL}, is a significant improvement in Recall ($+6.5\%$ absolute) and a slightly decrement in Precision (-0.8\% absolute), which give an absolute improvement in the F-measure of \tagme\ versus the one of \cite{witten} on short texts of about $+3\%$. As for \cite{soumen}, we recall that Chakrabarti et al used vectors over terms and over all involved senses (pages), conversely \tagme\ uses implicitly few short vectors, one vector per detected anchor and one dimension per detected sense (actually restricted to the un-discarded ones). This clearly induces a significantly better speed (more than one order of magnitude) and similar accuracy as commented in the following sections.

\smallskip As a final comment, we note that previous works \cite{cucerzan,soumen,wikify} deployed also the text surrounding anchors for boosting the efficacy of the disambiguation process. We tested these features in the design of \tagme\ but we either got worse accuracy or slower speed of annotation. We therefore dropped them from the final design of \tagme, and do not report these numbers for the lack of space. Nonetheless we believe that they could turn out to be useful in the applications of \tagme\ discussed in Sect. \ref{sec:concl}. So we plan to dig into them in the near future.

\subsection{Anchor pruning}
\label{sub:pruning}

The disambiguation phase produces a set of candidate annotations, one per anchor detected in the input text $T$. This set has to be {\em pruned} in order to possibly discard the un-meaningful annotations. These ``bad annotations'' are detected via a simple, yet effective, scoring function that takes into account only two features: the link probability $lp(a)$ of the anchor $a$ and the {\em coherence} between its candidate annotation $a \mapsto p_a$ (assigned by the Disambiguation Phase) and the candidate annotations of the other anchors in $T$. The effectiveness of the link probability in detecting significant anchors has been proved in \cite{witten}. The usefulness of the coherence was also shown in \cite{witten}, but limited to the case of {\em un-}ambiguous anchors. \tagme\ extends this notion to all anchors present in $T$ by introducing a novel formula that is based on the average relatedness between the candidate sense $p_a$ and the candidate senses $p_b$ assigned to all other anchors $b$. More precisely, if $S$ is the set of distinct senses assigned to the anchors of $T$ after the Disambiguation Phase (say $|S|>1$), we compute:

\[
 coherence(a \mapsto p_a) = \frac{1}{|S|-1} \sum_{p_b \in S \setminus \{p_a\}} {rel(p_b,p_a)}
\]

The goal of the pruning phase is to keep all anchors whose link probability is high or whose assigned
sense (page) is coherent with the senses (pages) assigned to the other anchors. We investigated five different
implementations of this idea: two are based on a proper arithmetic combination of the values $lp$ and
$coherence$; whereas the other three implementations deploy the classifiers C4.5, Bagged C4.5 and Support Vector Machine.
Each pruner computes for each candidate annotation a {\em pruning score}, say $\rho(a \mapsto p)$, and
then compares it against a properly set threshold $\rho_{\textsc{na}}$, so that
if $\rho(a \mapsto p) < \rho_{\textsc{na}}$ then that annotation for $a$ is discarded
by setting $a \mapsto \textsc{na}$.
The parameter $\rho_{\textsc{na}}$ allows to balance recall vs precision, and its impact
will be experimentally evaluated in Sect. \ref{sub:eval-all}.

The details of five pruners follow. The first two are very simple: one computes the
average of $lp$ and $coherence$ as $\rho_{{\tt AVG}}(a \mapsto p_a) = (lp(a) + coherence(a \mapsto p_a))/2$; the other
computes a linear combination $\rho_{{\tt LR}}(a \mapsto p_a) = \alpha \cdot lp(a) + \beta \cdot coherence(a \mapsto p_a) + \gamma$ in which the 3 parameters are trained via linear regression. Conversely the three
classifier-based pruners are implemented by taking $lp$ and $coherence$ as input features and return a value
(confidence) that can be interpreted as the ``probability of not-pruning'' the evaluated annotation $a \mapsto p$.

Sect.\ref{sub:eval-all} will evaluate the performance of these pruners and will show that, although
much simple in using just two features, they are all effective. Our final choice will be in favor
of $\rho_{{\tt AVG}}$ because of its simplicity (hence, speed) and its avoidance of any training step.

\section{Experimental evaluation}
\label{sec:evaluation}

In the following subsections we will address some key questions that pertain with the efficiency and efficacy of \tagme:

\begin{smallitemize}

\item How much is the coverage of Wikipedia's anchors in short texts like the ones occurring in web-search snippets and tweets? This is crucial, because we wish to understand how much useful can be the usage of Wikipedia anchors for annotation. (See Sect.~\ref{sub:WikiEval}.)

\item How much effective are the various disambiguation and pruning phases introduced above? And how do we choose the best algorithms and parameter settings for \tagme? (See Sect.~\ref{sub:tagmeAlone}.)

\item How does \tagme\ compare against best-known annotators \cite{soumen,witten} on short texts and on long texts ? (See Sect.~\ref{sub:shortText}-\ref{sub:longText}.)

\item How much fast is \tagme, and how its speed compares with the systems of \cite{soumen,witten}? (See Sect~\ref{sub:time}.)

\end{smallitemize}

\subsection{Coverage by Wikipedia anchors}
\label{sub:WikiEval}

First we want to evaluate the coverage of Wikipedia as catalog of senses for the annotation of
short texts drawn from the Web. We consider two types of text fragments: web snippets and
micro-blogging (namely, tweets), which constitute a {\em worst-case setting} for an annotator because
of their shortness and much poor textual composition. We derived these datasets by parsing about
5K tweets (of 14 words each, on average) and about 133K Web snippets (of 30 words each, on average)\footnote{Tweets and web snippets are gathered by using ``The 1000 most frequent web search queries issued to Yahoo! Search''. We randomly selected 300 queries from this dataset, performed searches on Tweeter and collected the first 20 results. For web snippets, we used almost all queries in that dataset and we collected the top 200 results from each query on Yahoo! search engine.}.

\begin{figure}
 \centering
 \includegraphics[width=\columnwidth]{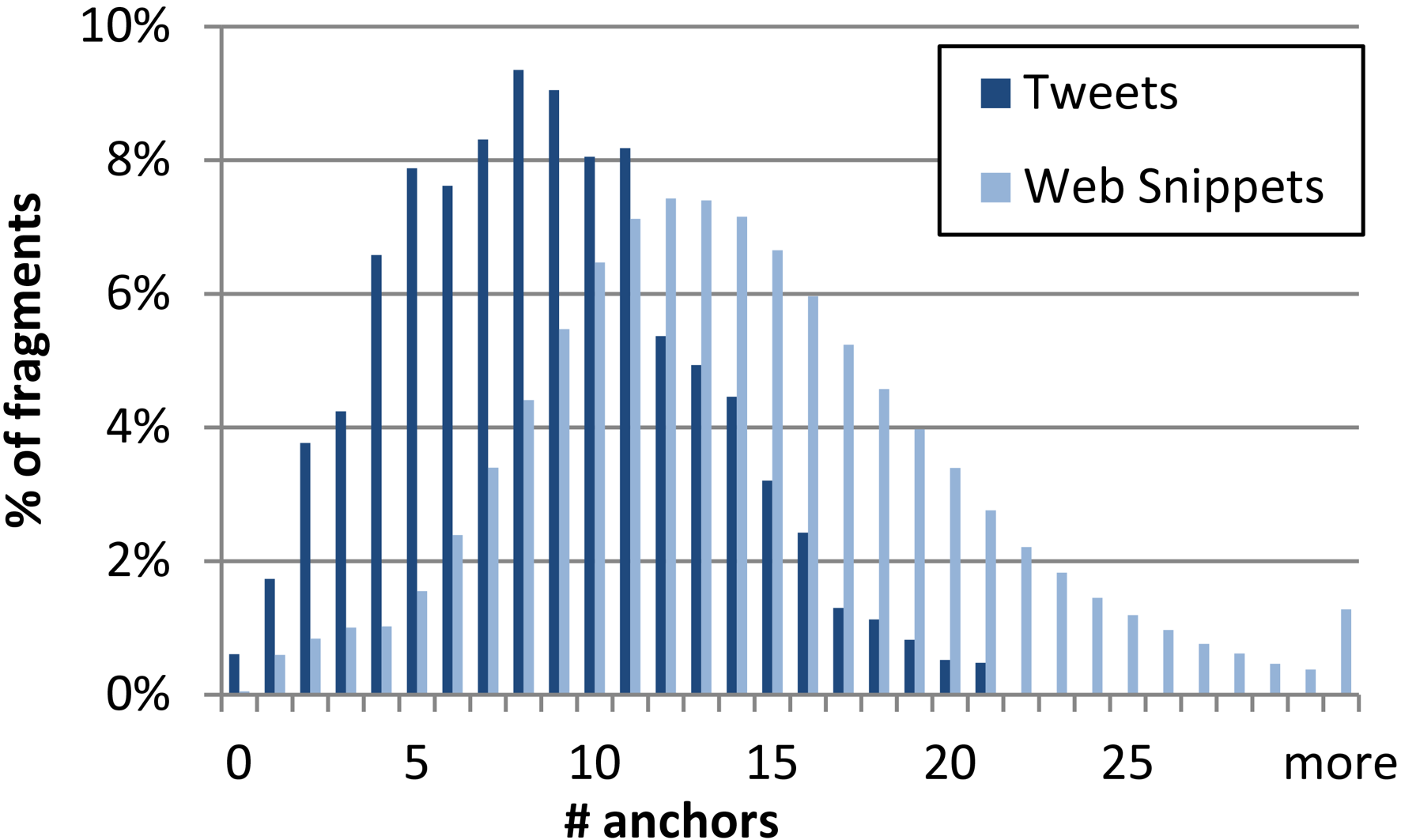}
 \caption{Number of Wikipedia anchors found in web snippets and tweets.}
 \label{fig:anchors-freq}
\end{figure}


Figure~\ref{fig:anchors-freq} reports statistics on the number of Wikipedia anchors found in those fragments: we point out that more than 93.9\% tweets and 98.5\% web-snippets have at least 3 anchors. So Wikipedia offers an unexpected large coverage of anchors/senses even in these challenging scenarios. We complemented this
quantitative test with a qualitative evaluation about the ``significance'' of the presence of an
anchor in a web-snippet or a tweet. We used $lp(a)$ as an estimate of the meaningfulness of $a$, as suggested
in \cite{wikify,witten}, and computed the distribution of anchor's link-probability among web-snippets and tweets.
Results are not reported for the lack of space, however we note that for at least $95\%$ of the short
texts (both snippets and tweets) the top-lp is larger than $6.5$\%, which \cite{witten} considered a strong
indication of a significant anchor. This percentage remains high, namely $>90\%$, when we average
the values of the top-5 lps of the anchors detected in the short text. These results support our hypothesis that
Wikipedia is a significant catalog of senses also for the short texts drawn from the Web.

\subsection{Setting up TAGME}
\label{sub:tagmeAlone}

For setting up \tagme\ we used three datasets derived from Wikipedia, as done in \cite{witten}. The first dataset, denoted \WikiDis, consists of 1.4M short fragments randomly selected from Wikipedia pages. Each fragment consists of about 30 words (like Web snippets' composition). To avoid any advantage to \tagme, we were careful in selecting fragments that contain at least one ambiguous anchor-text (i.e. $|\Pg(a)| > 1$). The second dataset, denoted \WikiAnno, consists of 150K fragments constructed as follows. Since Wikipedia authors usually link only the first occurrence of an anchor $a$ in a Wikipedia page $z$, a short fragment $T_z$ randomly drawn from $z$ could contain occurrences of $a$ which are un-annotated. Therefore we expand \WikiAnno\ by extending the annotation $a\mapsto p$ occurring in $z$ to all occurrences of $a$ in this page (and thus to the ones in the fragment $T_z$). After this expansion, \WikiAnno\ contains about 1.5M anchor occurrences, of which 47\% are annotated. The third dataset, denoted \WikiLong, consists of about 10K randomly selected Wikipedia articles that contain at least 10 links. This dataset contains about 270K links in total, and models the case of highly linked and long texts.

To evaluate the performance of the Disambiguation Phase, we use standard precision and recall scores; whereas for the overall annotation process (disambiguation+pruning) we follow \cite{soumen} and thus focus
on the precision $P_{ann}$ and recall $R_{ann}$ measures that are computed on the set of anchors
which are annotated in the ground truth (i.e. the corpora above). These last measures are much
demanding because they ask for a perfect match between the annotation in the ground truth and the one
obtained by the tested system. If the goal is to identify {\em topics} in the text fragment, then it
doesn't matter which anchors got annotated but which senses got linked. So, let ${\cal G}(T)$ be the
senses (pages) associated to the anchors of $T$ in the ground truth, and let ${\cal S}(T)$ be the senses
identified by the tested system over $T$. As in \cite{witten}, we define a topic-based notion of precision ($P_{topics}$) and recall ($R_{topics}$) over ${\cal G}(T)$ and ${\cal S}(T)$.

\subsubsection{Setting the disambiguation phase}
\label{sub:eval-disamb}

We recall that \dc\ and \dt\ are the two approaches to disambiguation proposed in Sect.~\ref{sub:disambig}. Here we experiment them by splitting \WikiDis\ into two parts: one contains 400K anchors and is used for training, the other contains the remaining 1M anchors.

\smallskip \noindent {\bf Approach \dc.} We trained a C4.5 classifier that was shown to achieve the best results for disambiguation in \cite{witten}. We also tested several values of the parameter $\tau$ (which controls the pruning {\em ex ante} of some pages), and discovered that $\tau=0.5$\% gives the best results: larger values do not gain precision, but reduce significantly the recall.

\smallskip \noindent {\bf Approach \dt.} In addition to $\tau$, this approach depends on a parameter $\epsilon$ which controls the ``roughness'' in the value of the voting score. Setting $\epsilon$ close to zero leads \dt\ to always select the sense that achieves the highest value of $rel_a$ (i.e. the most related sense, shortly {\tt MR}), while $\epsilon$ close to 1 leads \dt\ to select the sense with the highest commonness (i.e. the most common sense, shortly {\tt MC}). If all senses $p_a$ get $rel_a(p_a)=0$, we decided to set $a\mapsto {\textsc{na}}$ because it is reasonable to argue that they are un-related with the topics of the input text.

\begin{figure}[h]
  \includegraphics[width=\columnwidth]{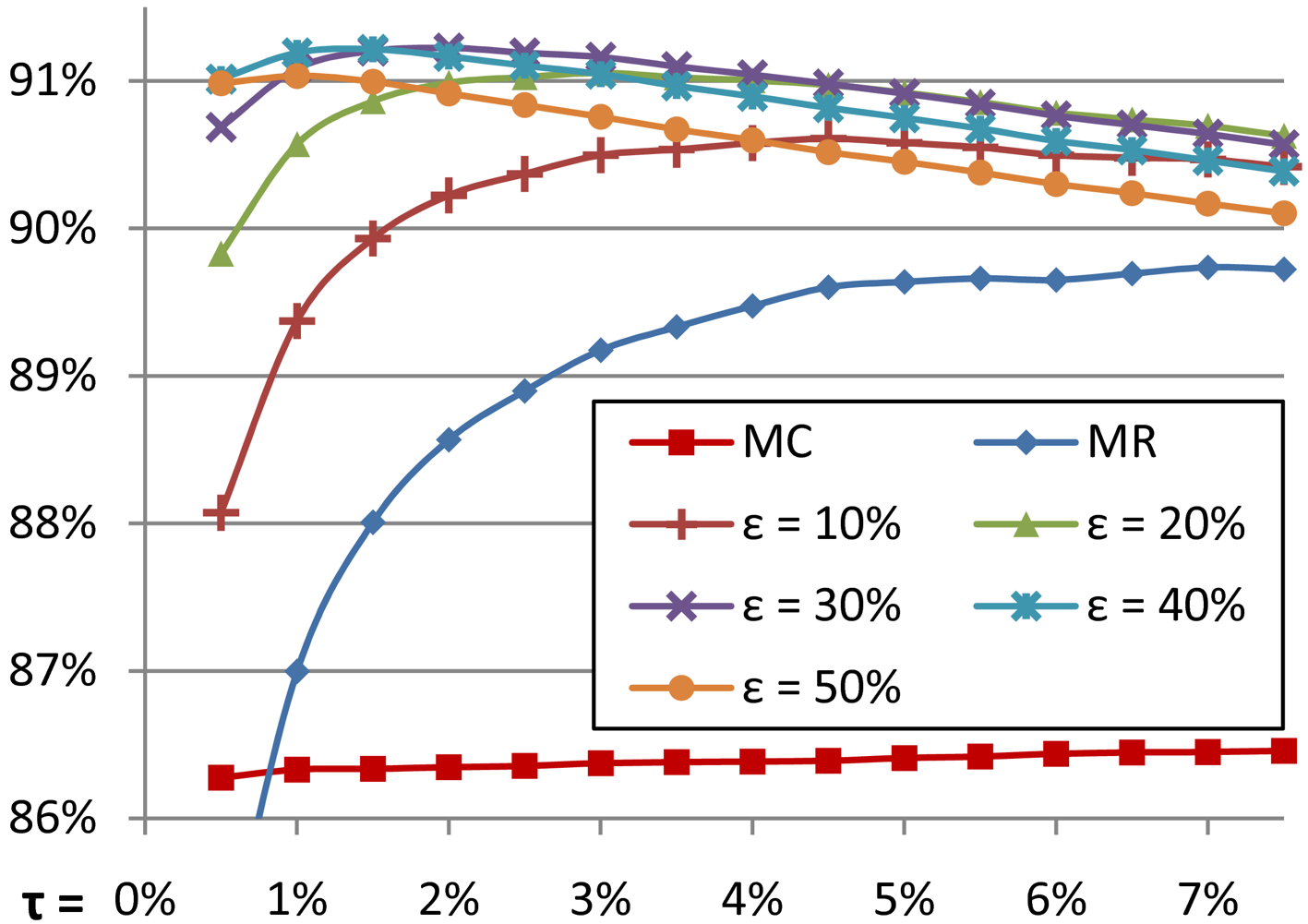}%
  \caption{Performance of \dt\ by varying $\epsilon,\tau$ over the training dataset, where {\tt MC} and {\tt MR} denote the choice of the Most Common (ie. $\epsilon=100\%$) and the Most Related (ie. $\epsilon=0\%$) sense, respectively.}%
  \label{fig:train-dt}%
\end{figure}

Figure~\ref{fig:train-dt} plots the F-measure for \dt\ by varying $\tau$ and $\epsilon$. The performance of \dt\ for values of $\epsilon > 50$\% are not plotted because they are very close to the \emph{most-common sense} {\tt MC}. Overall, lower values of $\epsilon$ are better; experiments lead us to choose $\tau = 2\%$ and $\epsilon = 30\%$.

\smallskip Given these parameter settings, we compared \dc\ and \dt\ over the testset of 1M anchors drawn from \WikiDis. Performances are shown in Table~\ref{tab:disamb} below and, although precision and recall are very close, we decided to use \dt\ in \tagme\ because of three main reasons: (i) it has a better F-measure, (ii) the choice of $\tau=2\%$ discards many un-significant pages and thus gains much speed, as explained in Sect.~\ref{sub:disambig}, (iii) \dt\ depends on a threshold $\epsilon$ that gives flexibility: we can increase $\epsilon$ if the input texts are too ambiguous (and thus choose more often the most-common sense) or decrease $\epsilon$ if the input texts are more focused (and thus choose more often the most-related sense).

\begin{table}[h]
  \begin{center}
	  \begin{tabular}{lccc}
		  \hline
			& Precision & Recall & F-measure \\
			\hline
			\dc\ & \textbf{91.7} & 89.9 & 90.8 \\
			\dt\ & 91.5 & \textbf{90.9} & \textbf{91.2} \\
			\hline
		\end{tabular}	
	\end{center}
	\caption{Disambiguation performance over {\normalfont {\WikiDis}}.}
\label{tab:disamb}
\end{table}

\subsubsection{Setting the pruning phase}
\label{sub:eval-all}

As detailed in Sect.~\ref{sub:pruning}, the pruning step hinges on two features, $lp$ and $coherence$. We tested two combination of these features (AVG and LR) and three different classifiers deploying these features (C4.5, Bagged C4.5 and SVM). We trained these classifiers over 50K short texts extracted from \WikiAnno, and then tested them over the remaining 100K short texts. The problem we faced was to generate both positive and negative training cases from the 50K texts. We thus proceeded as follows. We run the disambiguator \dt\ over those texts and compared its annotation with the (available) ground truth for them, thus totalling $\approx 10 \times 50K = 500K$ anchor annotations. There are three cases: if the anchor is linked in the ground truth and the linked page coincides with the one assigned by \dt, then it is a positive case (with its $lp$ and $coherence$ values); if it is linked in the ground truth but the linked page differs from the one assigned by \dt, then it is discarded from the training; all other cases are considered as negative cases for the training set (with their $lp$ and $coherence$ values). At the end remained a total of $460$K training cases. Moreover, in order to train the three parameters of the approach based on linear regression (LR), we transformed the boolean class of the ground truth (linked or not linked) into a numeric value: we set $\rho_{\tt LR} = 1$ for positive cases (i.e. linked anchors) and $\rho_{\tt LR} = 0$ for negative cases.

\begin{table}%
  \begin{center}
	  \begin{tabular}{lccc}
		  \hline
			& $P_{ann}$ & $R_{ann}$ & F-measure \\
			\hline
			Only $lp$ & 75.50 & 72.01 & 73.71\\
			AVG & 76.27 & 76.08 & 76.17\\
			LR & 76.49 & 75.74 & 76.10\\
			C4.5 &  {\bf 76.72} & {\bf76.22} & {\bf 76.47}\\
			Bagged C4.5 & 76.54 & {\bf 76.22} & 76.38\\
			SVM & 76.25 & 75.96 & 76.11\\
			\hline
		  \hline
			& $P_{topics}$ & $R_{topics}$ & F-measure \\
			\hline
			Only $lp$ & 76.85 & 76.65 & 76.75\\
			AVG & 78.41 & 77.48 & 77.94\\
			LR & 78.42 & 77.03 & 77.72\\
			C4.5 & 76.78 & {\bf 79.69} & {\bf 78.21}\\
			Bagged C4.5 & {\bf 79.13} & 77.12 & 78.11\\
			SVM & 78.91 & 77.13 & 78.01\\
			\hline
		\end{tabular}	
	\end{center}
	\caption{Performance of various pruners over {\normalfont \WikiAnno}, using \emph{annotation} and \emph{topics} metrics.}
\label{tab:pruning}
\end{table}

After training, we evaluated our pruners over the remaining 100K fragments of \WikiAnno\ by varying $\rho_{\textsc{na}}$ in $[0,1]$ using a step of $0.01$ (recall that $\rho_{\textsc{na}}$ controls the {\em sensibility} of our annotation process). In these experiments we included another simple pruner that we called ``Only $lp$'' which uses only the link probability of the evaluated anchor to prune the un-meaningful annotations. By comparing this approach against the others, we can evaluate the significance of using the feature \emph{coherence} in addition to the link-probability in the pruning step. We tested also the case of {\em coherence-only} feature but performance was worse, and is not reported for space reasons.

Table~\ref{tab:pruning} summarizes all experimental results for the settings of $\rho_{\textsc{na}}$ that yield the highest F-measure, using 2-fold cross validation. As expected, \emph{annotation} measures are more severe than \emph{topics} measures, although there are dependencies between them, and indeed the ranking of the pruners is the same in both of them. The overall performance of all pruning approaches is very close to each other. Results also show that ``Only $lp$'' is surpassed by all other approaches that deploy also \emph{coherence}, which confirms the usefulness of both features in the pruning phase. As a result, and inspired by the {\em Occam Razor} principle, we decided to implement in \tagme\ the simplest pruning method based on  $\rho_{{\tt AVG}}$. This is because SVM is very slow, the others are as fast as AVG but they need a training step that we prefer to avoid in the Web context. The final setting for \tagme\ uses $\rho_{\textsc{na}} = 0.2$, however the on-line version of \tagme\ offers the possibility to modify this value.

\subsection{Comparing annotators on short texts}
\label{sub:shortText}

We compare the best \tagme\ setting (namely \dt\ plus ${\tt AVG}$) against the Milne\&Witten's system re-built (for fairness) over the same Wikipedia snapshot used by \tagme. Since we could not get access to Chakrabarti's system\footnote{Chakrabarti's personal communication.}, we do not include it in this comparison. In any case, we recall that this system cannot be used in our context because it is very slow: it takes $>2$ secs per $15$ anchors \cite{soumen}. This is more than one order of magnitude slower than \tagme, which takes less than 2ms per anchor (see next Sect. \ref{sub:time} for details). Nevertheless, Chakrabarti's system will be considered when annotating long texts because of the results reported in \cite{soumen}, see next Sect. \ref{sub:longText}.

\smallskip The first experiment compares the disambiguation phase of \tagme\ against the one of Milne\&Witten's system over the testset of \WikiDis. To make the comparison wider, we considered also two other (simpler) disambiguators: one selects always the most-common sense from $\Pg(a)$ (i.e. statistically-driven choice), and the other randomly selects a page from $\Pg(a)$ (i.e. oblivious choice).

\begin{table}[h]
  \begin{center}
	  \begin{tabular}{lccc}
		  \hline
			& Precision & Recall & F-measure \\
			\hline
			Random & 32.2 & 32.2 & 32.2\\
			Most Common & 85.8 & 86.8 & 86.3 \\
			Milne\&Witten & \textbf{92.3} & 84.6 & 88.3\\
			\dt\ (in \tagme) & 91.5 & \textbf{90.9} & \textbf{91.2} \\
			\hline
		\end{tabular}	
	\end{center}
	\caption{Performance of various disambiguation algorithms over the short texts of {\normalfont \WikiDis}.}
\label{tab:disambALL}
\end{table}

Results are shown in Table~\ref{tab:disambALL}. With respect to Milne\&Witten, our disambiguator \dt\ yields a significant improvement in Recall ($+6.5\%$ absolute) and a slightly decrement in Precision (-0.8\% absolute), for an overall absolute improvement in F-measure of about $3\%$. This is due to our voting scheme that deploys the relatedness among the senses associated to all anchors in the short input text, and thus not only among the senses of the un-ambiguous anchors (which are possibly absent in short texts, as commented in Sect.~\ref{sub:disambig}).

\smallskip As a final check for fairness, we also compared these results against the ones presented by Milne\-\&\-Witten in \cite{witten}. In that paper, the authors evaluated their system over a collection of $100$ full-articles of Wikipedia, each containing at least $50$ links (for a total amount of $11$K anchors). Their system yielded an overall F-measure on disambiguation of about $96\%$, which is larger than the $88\%$ reported in Table~\ref{tab:disambALL}. The reason is that our texts are short and more ambiguous, and thus more difficult to be disambiguated: in fact, on our datasets the choice of a random sense gets $F\approx 32\%$ and the choice of the most-common sense gets $F\approx 86\%$; whereas on Milne\&Witten's dataset, these numbers were $53\%$ and $90\%$ respectively. This remarks further that the performance of \tagme's disambiguation is much effective.

\begin{table}[h]
  \begin{center}
	  \begin{tabular}{lccc}
		  \hline
			& $P_{ann}$ & $R_{ann}$ & F-measure \\
			\hline
			Milne\&Witten & 69.32 & 69.52 & 69.42\\
			\tagme\ & \textbf{76.27} & \textbf{76.08} & \textbf{76.17}\\
			\hline
		  \hline
			& $P_{topics}$ & $R_{topics}$ & F-measure \\
			\hline
			Milne\&Witten & 69.60 & 69.80 & 69.70\\
			\tagme\ & \textbf{78.41} & \textbf{77.48} & \textbf{77.94}\\
			\hline
		\end{tabular}	
	\end{center}
	\caption{Performance of annotators on short texts.}
\label{tab:shortALL}
\end{table}

\smallskip Finally, we compared the overall annotation process in the two systems by using the testset of \WikiAnno. Results are reported in Table~\ref{tab:shortALL}, where we notice that \tagme\ significantly improves Milne\&Witten's system over short texts both in precision and recall, by about $8-9\%$ absolute. The reason is that many features used by Milne\&Witten's system are not effective on short texts: indeed, they considered features like location and frequency of anchors (which may be ``undefined'' or even misleading on short texts), as well as they considered only the un-ambiguous anchors to compute a coherence-score (and these are often absent in short texts, as we commented above).

\begin{figure}[h]
\includegraphics[width=\columnwidth]{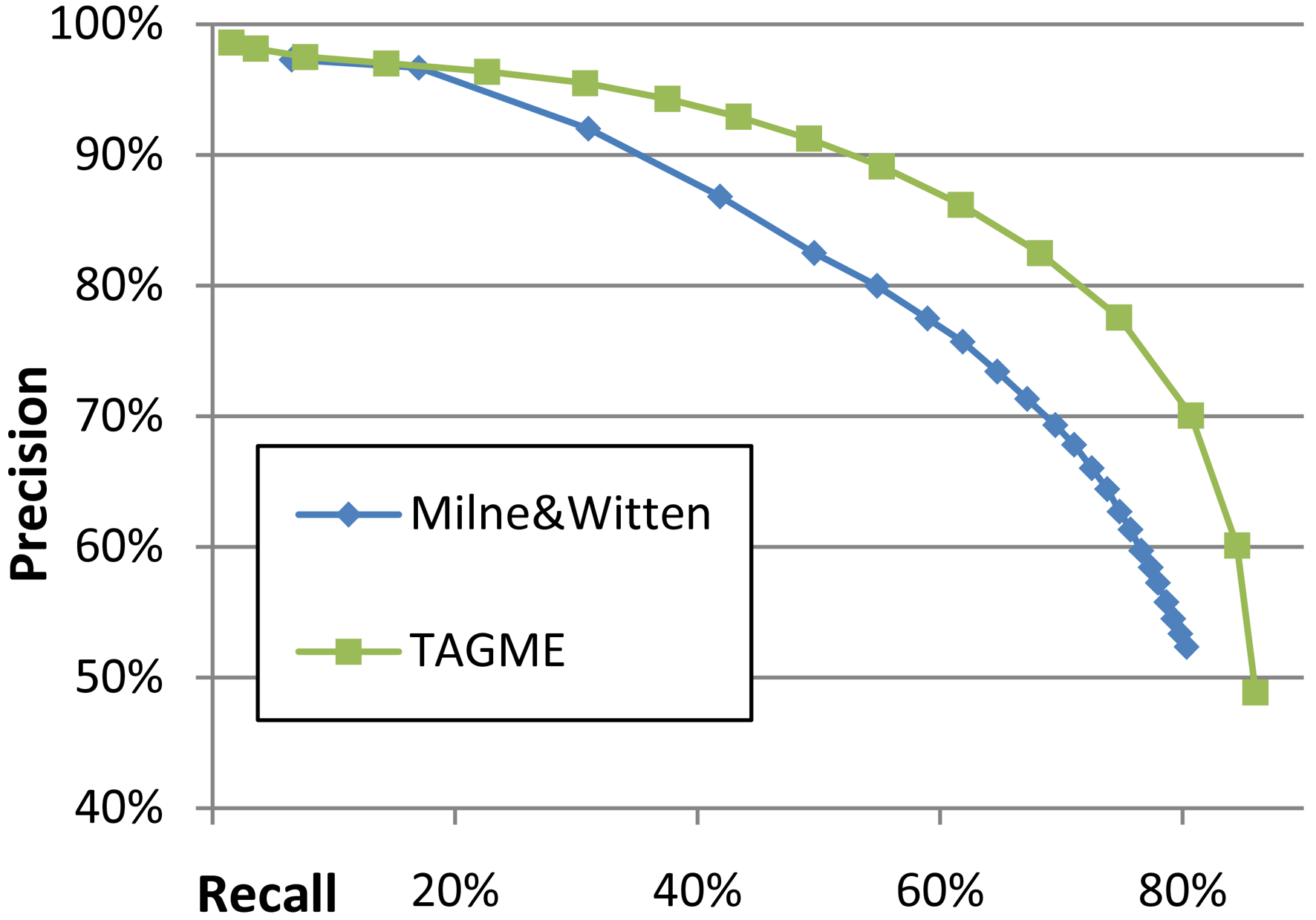}%
\caption{Performance of annotators on short texts, by varying the value of $\rho_{\normalfont \textsc{na}}$.}%
\label{fig:overall-short}%
\end{figure}

Since also the system by Milne\&Witten offers the possibility to balance precision vs recall, we analyzed the impact of $\rho_{\textsc{na}}$ on the performance of the two annotators. Figure~\ref{fig:overall-short} reports the comparison which shows that \tagme\ improves clearly the other approach for almost all values of $\rho_{\textsc{na}}$.

\subsection{Comparing annotators on long texts}
\label{sub:longText}

If the input text is long, we do not want to change \tagme's architecture because we want to obtain a software whose time complexity scales {\em linearly} with the number of anchors in the input text (see below). So we shift a text window of about 10 anchors over the long input text, and apply \tagme\ on each window in an incremental way. It is clear that this approach gives advantage to both Chakrabarti's and Witten\&Milne's systems in terms of precision/recall of the annotation, because they deploy the full input text (and thus probably more than 10 anchors). Nevertheless, we decided to stick on this unfavorable setting for \tagme\ in order to stress its performance.

\begin{figure}[t]
\includegraphics[width=\columnwidth]{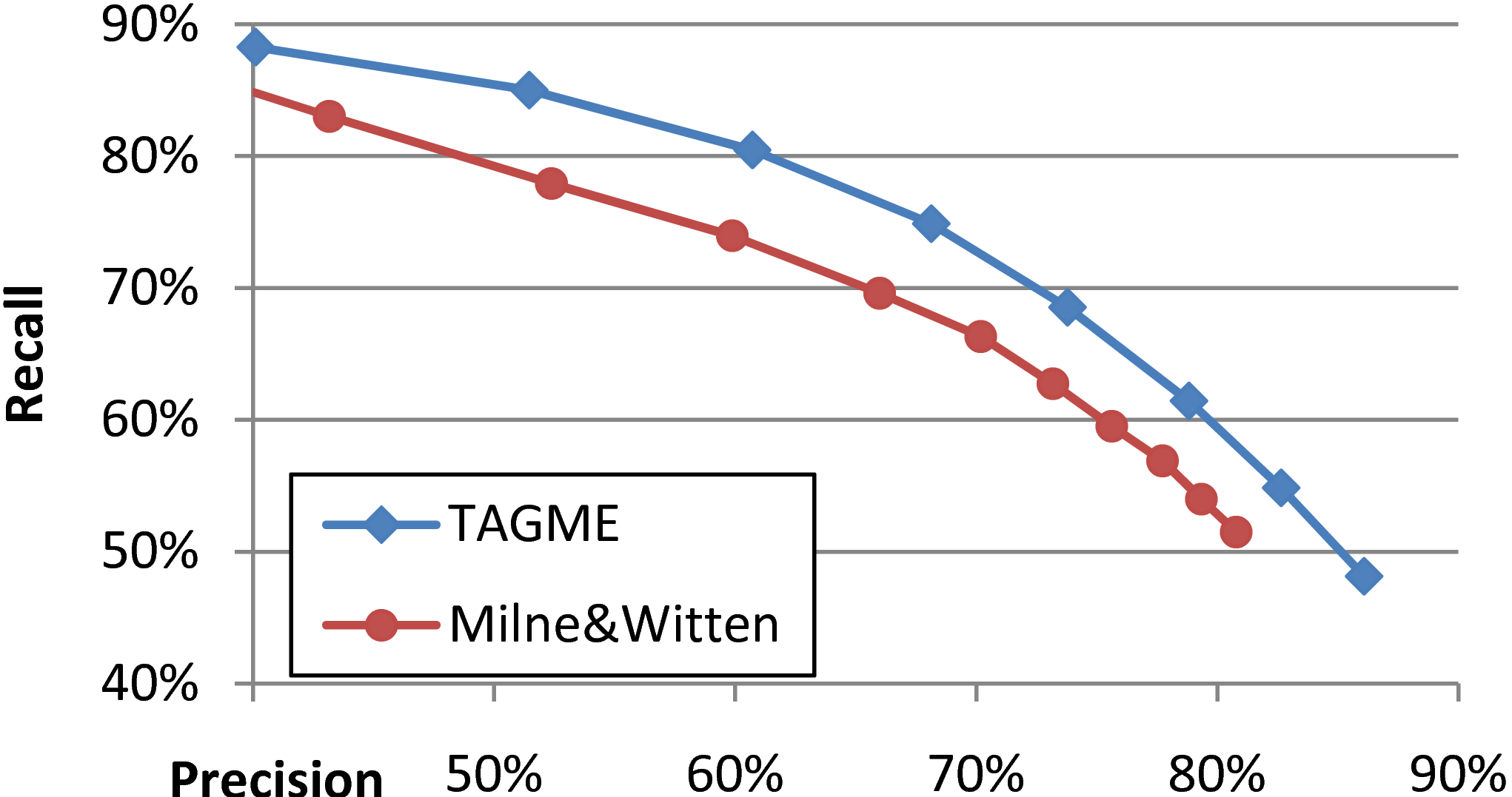}
\caption{Performance of annotators on the long texts of \WikiLong.}
\label{fig:overall-long}
\end{figure}

Our first experiment is on our dataset \WikiLong\ (see Sect. \ref{sub:tagmeAlone}), and compared the only two available annotators: \tagme\ and Milne\&Witten's system. Figure~\ref{fig:overall-long} reports the precision/recall curves as the value of $\rho_{\textsc{na}}$ varies. Surprisingly \tagme\ improves Milne\&Witten's system uniformly when texts have from ten to hundreds of anchors to be annotated. Given these results we decided to dig into \WikiLong\ in order to evaluate  the performance of the two systems as a function of the number of anchors in the long input text. For space reasons we cannot plot these results but we briefly state that, as expected, as this number grows the performance of Milne\&Witten's system improves (and approaches an F-measure of about 74\%, as stated in \cite{witten} for long and many-linked documents) whereas the performance of \tagme\ drops (and approaches 72\% from the 78\% achieved on short texts). It goes without saying that \tagme\ is designed for short texts, and its parameter settings have not been re-trained for this long-text case. We plan to investigate deeply the case of long input-texts for \tagme\ in the near future, possibly designing a robust variant that dynamically adapts its settings based on the length of the input text to be annotated.

\begin{figure}[h]
 \centering
 \includegraphics[width=\columnwidth]{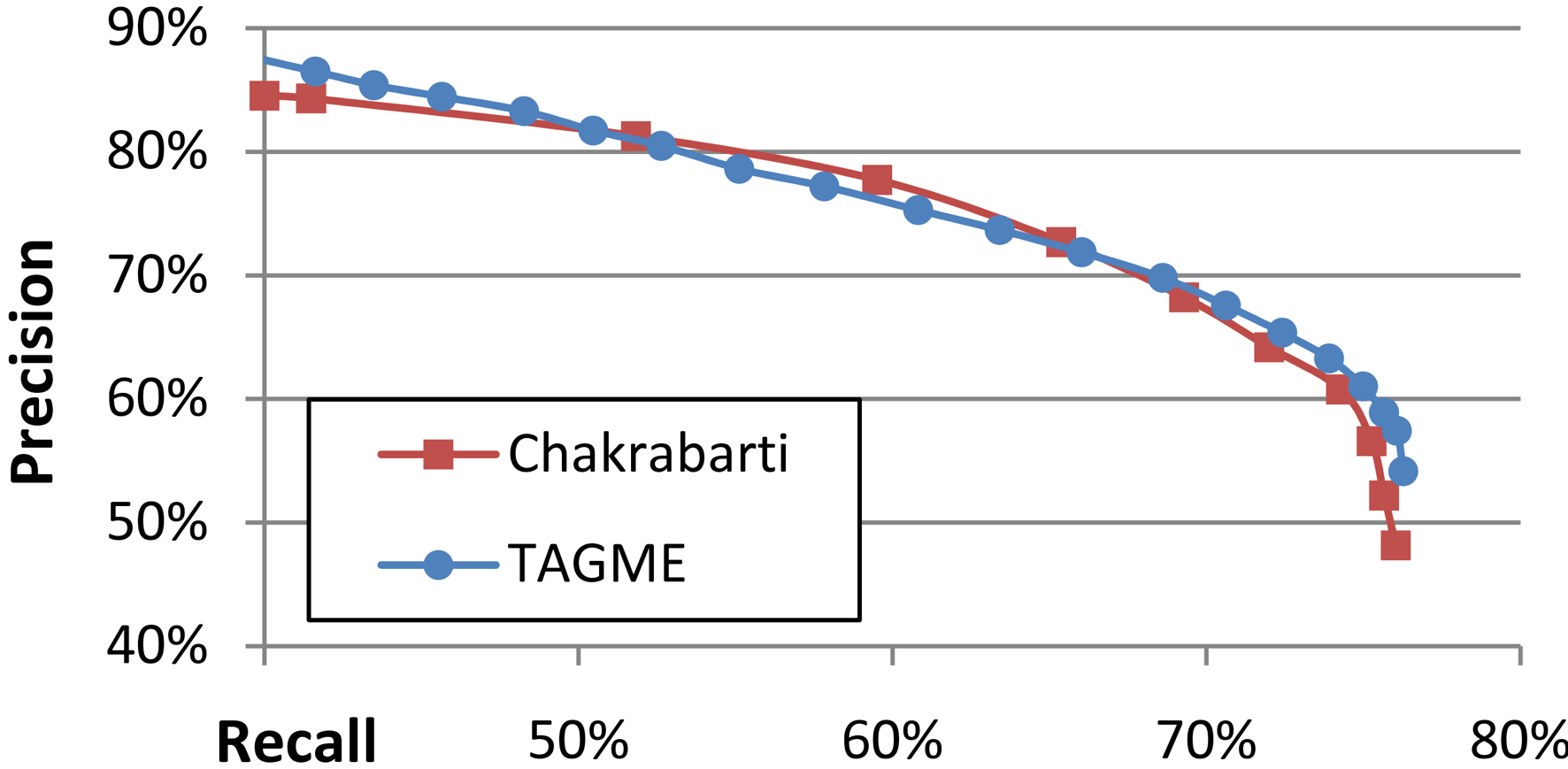}
 \caption{Performance of Chakrabarti's annotator and {\normalfont \tagme} over the {\normalfont \textsc{iitb}} dataset.}
 \label{fig:iitb}
\end{figure}

Our second experiment is aimed at deriving some information about the Chakrabarti's system. So, we downloaded their {\sc iitb} dataset and run \tagme\ over it. This way we can use the performance figures reported in \cite{soumen} to compare all three known systems. Figure~\ref{fig:iitb} reports only the performance of Chakrabarti's and \tagme's systems because, as reported in \cite{soumen}, Milne\&Witten's system performs so poorly on this dataset that its recall/precision curve is far to the left and thus is dropped to ease the reading of the figure. It is interesting to observe that, even with the severe limitation imposed by the shift-based approach, \tagme\ is competitive to Chakrabarti's system in terms of precision/recall figures, with the advantage of being more than one order of magnitude faster (see below).

\subsection{On the time efficiency of TAGME}
\label{sub:time}

The most time consuming step in \tagme's annotation is the calculation of the relatedness score of Sect. \ref{sub:disambig}, because anchor detection and other scores require time {\em linear} in the length of the input text $T$. If $n$ is the number of anchors detected in $T$, $s$ is the average number of senses potentially associated with each anchor, and $d_{in}$ is the average in-degree of a Wikipedia page, then the time complexity of the overall annotation process is $O(d_{in} \times (n\times s)^2)$. On our datasets of short texts it is $n \approx 10$, $s \approx 5$ and $d_{in} \approx 50$, so that our current implementation of \tagme\ takes 1.7ms per anchor and about 18ms per short text on a commodity PC\footnote{In detail, anchor parsing takes about 6.2ms, disambiguation and pruning about 12.5ms per input text. Of course, code engineering may speed-up \tagme, and this will be addressed in the future.}. This is more than one order of magnitude faster than the time performance reported in \cite{soumen}. Additionally, when \tagme\ is applied on long texts of $L$ anchors (possibly $L \gg 10$), it slides and processes a window of about $w=10$ anchors over the input text. This way, \tagme\ can re-compute incrementally the scores and thus pay $W_{cost} = O(d_{in} \times w\times s^2)$ time per window. This is $O(L \times W_{cost})$ time in total, which results linear in the number of anchors to be annotated. Conversely, Chakrabarti's system scales ``mildly quadratically'' in $L$, as stated in \cite{soumen}.

\begin{figure}[h]
\includegraphics[width=\columnwidth]{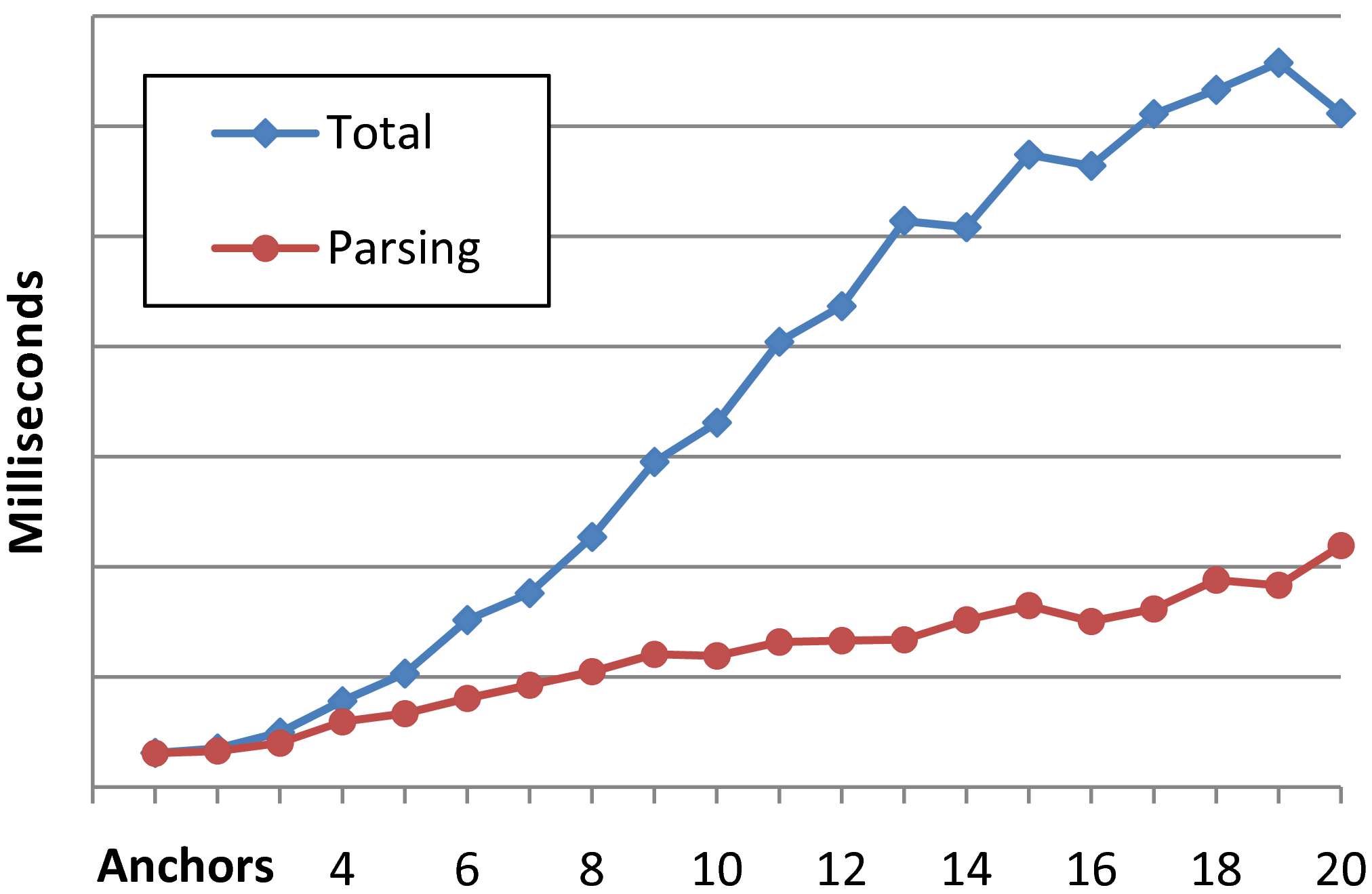}%
\caption{Time performance of {\normalfont \tagme}.}%
\label{fig:time}%
\end{figure}

Figure~\ref{fig:time} shows the time performance of \tagme\ as the number of anchors in the input text grows\footnote{Text fragments are drawn from \WikiAnno.}. The line ``Total'' indicates the average time taken by \tagme\ for the complete annotation process (i.e. parsing + disambiguation + pruning). When the input text has less than 10 anchors (i.e. it is short), the trend is roughly quadratic as predicted by the above analysis with a significant cost induced by the parsing step. As the number of anchors to be annotated grows, the time tends to grow linearly (as predicted) with a parsing cost which becomes almost negligible.
We didn't report the time performance of Milne\&Witten's system because it is significantly slower than \tagme: it takes about 95ms per text on average, and this would have jeopardized the reading of the figure. Let it be said that Milne\&Witten's software allows to set up a {\em cache} in order to speed up the annotation process. We tried this setting too, but it incurred in a large internal-memory allocation of about 1.4GB for the initialization step, and then its Java heap-space overflowed the 2GB available on our PC just after the annotation of few thousands of short texts (i.e. 6-7K). Anyway, the performance obtained with the help of the cache (which is 18.36ms on avg per short text) is comparable with \tagme, but \tagme\ uses just 200MB of internal memory independently of the number of processed texts.

\section{Conclusion and future works}
\label{sec:concl}

In the light of the experiments conducted on Wikipedia-based datasets, one could reasonably ask: does \tagme\ achieve the same effective performance {\em in the wild}? There are two issues that let us argue positively about this: (1) the {\sc iitb} dataset is a manually annotated set of news stories drawn from the Web, and there \tagme\ is superior either in precision/recall or speed to the state-of-the-art systems (see Sect. \ref{sub:longText}); (2) the user-study conducted in \cite{witten} confirmed that performance yielded over large datasets drawn from Wikipedia are good predictors of annotation performance {\em in the wild}, and indeed our datasets were larger and variegate than the ones used in that paper. In addition to these two positive witnesses, we are currently setting up a much larger user-study over Mechanical Turk\footnote{{\tt http://www.mturk.com}} with the twofold goal of creating a manually-annotated dataset much larger than the one offered by \cite{soumen} and extending the tests of \tagme.

\smallskip We believe that \tagme, like the systems of \cite{soumen,witten}, has implications which go far beyond the enrichment of a text with explanatory links. We are currently investigating the impact of \tagme's annotation onto the performance of our past system {\sc SnakeT} \cite{snaket} for the on-the-fly labeled clustering of search-engine results (see also {\sc Clusty.com} or {\sc Carrot}). In fact {\sc SnakeT}, as most of its competitors (see e.g. \cite{clustering-engines}), is based only on syntactic and statistical features and thus it could benefit from \tagme's annotation to improve the effectiveness of the labeling and the clustering phases. Furthermore, we are studying how other by-products of Wikipedia--- such as {\tt DBpedia.org}, {\tt Freebase.com}, Kylin \cite{kylin} or YAGO \cite{yago}--- could be used in \tagme\ to better relate and/or assign senses to text anchors.

Finally, we plan to investigate the application of \tagme\ in Web Advertising: the explanatory links and the structured knowledge attached to plain-texts could allow the efficient and effective resolution of ambiguity and polysemy issues which often occur when advertiser's keywords are matched against the content of Web pages offering display-ads.



\end{document}